\documentclass{ws-procs9x6}

\newcommand{\refeq}[1]{(\ref{#1})}
\def\etal {{\it et al.}}

\begin{document}

\title{LORENTZ VIOLATION IN A \\
UNIFORM GRAVITATIONAL FIELD}

\author{Y.\ BONDER}

\address{Physics Department, Indiana University\\
Bloomington, IN 47405, USA\\
E-mail: ybonder@indiana.edu\\
(Dated: IUHET 579, July 2013)}

\begin{abstract}
We present a method to calculate the nonrelativistic hamiltonian for the minimal Standard-Model Extension matter sector in a uniform gravitational field. The resulting hamiltonian coincides with earlier results in the corresponding limits but it also includes spin-dependent terms that were previously unknown. The phenomenology associated with this hamiltonian is briefly discussed.  
\end{abstract}

\bodymatter

\phantom{x}
\vskip 1pt
\noindent
Lorentz invariance lies at the core of our current conception of physics. Thus, it seems necessary to test empirically Lorentz invariance. The Standard-Model Extension (SME) is a framework that includes the Standard Model and General Relativity plus all possible Lorentz violating terms,\cite{SME, Kost} and it is widely used as a guide to search and parametrize Lorentz violation. As was pointed out by Kosteleck\'y,\cite{Kost} gravity couplings allow access to some SME coefficients that are otherwise unobservable. Thus, to explore all possible ways Lorentz violation could manifest, one needs to study the matter-gravity SME sector.

Kosteleck\'y and Tasson\cite{KostTasson} studied the matter-gravity SME sector and their analysis led to bounds on some previously unmeasured SME coefficients.\cite{DataTables} However, in Ref.\ \refcite{KostTasson} they do not consider the SME coefficients associated with spin. Our proposal is to study the SME matter-gravity sector with a different --- and complementary --- approach. This allows us to obtain the nonrelativistic hamiltonian for the free Dirac fermion minimal SME sector in a uniform newtonian gravitational potential, including spin effects.

We now outline our work hypothesis. The fermions are assumed to be test particles, namely, the spacetime curvature generated by these particles is neglected. Moreover, the background spacetime is chosen to be the flat (and torsion-free) spacetime but as seen by a uniformly accelerated observer. We also assume that the observer's acceleration can be identified with a uniform newtonian gravitational potential $\Phi$. Note that, in the situation at hand, the no-go theorem precluding us from having explicit Lorentz violations in curved spacetimes\cite{Kost} does not apply, and it is consistent to consider explicit Lorentz violations. This allows us to ignore the mechanisms from which the Lorentz violations would spontaneously emerge, simplifying considerably our analysis. Furthermore, we require the SME coefficients to have vanishing covariant derivatives, which is equivalent to the condition that the coefficients are constant for an inertial observer.

To construct the hamiltonian we use, as a starting point, the action of the corresponding SME sector in a curved background spacetime.\cite{Kost} We work in Fermi-like coordinates associated with a uniformly-accelerated and nonrotating observer, as done in Ref.\ \refcite{Hehl}. In principle, the equations of motion can be used to read off a hamiltonian, defined as the generator of time translations. However, one must invert a matrix contracted with the wavefunction time derivative. To invert this matrix we use the field redefinition method described in Ref.\ \refcite{KostTasson} that, in addition, ensures that the resulting hamiltonian is hermitian with respect to the standard inner product of nonrelativistic quantum mechanics.

We are interested in tabletop experiments where the particles are nonrelativistic. Therefore, we seek the nonrelativistic hamiltonian to first order in the SME coefficients. To do so, we perform three Foldy-Wouthuysen transformations\cite{FW} that decouple the particle and antiparticle degrees of freedom. We also remove all the unphysical terms, namely, those that can be canceled with unitary transformations. For simplicity we only present here the nonrelativistic hamiltonian for the particles (as opposed to the antiparticles) to first order in $\Phi$;  the resulting hamiltonian is
\begin{eqnarray}
H_{\rm NR}&=&A(1+\Phi)+B_i \sigma_i(1+\Phi) +\frac{1}{m}C_i \left(p_i+\frac{1}{2} \Phi p_i +\frac{1}{2}p_i \Phi \right) 
\nonumber\\
&&+\frac{1}{m} D_{ij}\sigma_j \left(p_i+\frac{1}{2} \Phi p_i +\frac{1}{2}p_i \Phi \right)+\frac{1}{2m} E_{ij} p_i (1+\Phi) p_j
\nonumber\\
&&+\frac{1}{2m^2} F_{ijk} \sigma_k \left( p_i p_j+\frac{3}{2}p_i\Phi p_j +\frac{3}{2} p_j\Phi p_i \right)
\nonumber\\
&&+\frac{1}{m^2}G_{ijk} \sigma_k\left(p_i p_j+\frac{1}{2} p_i \Phi p_j + \frac{1}{2} p_j \Phi p_i \right),\label{bonder:Hamiltonian}
\end{eqnarray}
where
\begin{eqnarray}
A&=&m+\hat{a}_0-m\hat{c}_{00} -m \hat{e}_0,
\end{eqnarray}
\begin{eqnarray}
B_i&=&-b_i+m \hat{d}_{i0}+\frac{1}{2}\epsilon_{ijk}\left(-m\hat{g}_{jk0}+H_{jk} \right) ,
\end{eqnarray}
\begin{eqnarray}
C_i&=&a_i -m\hat{c}_{0i}-m\hat{c}_{i0}-m e_i,
\end{eqnarray}
\begin{eqnarray}
D_{ij}&=&-\delta_{ij}\hat{b}_0 +m \delta_{ij}\hat{d}_{00} +md_{ij} +\epsilon_{ijk}\left(\frac{1}{2}  (\partial_k\Phi) +m \hat{g}_{0k0} - \hat{H}_{0k} \right),
\end{eqnarray}
\begin{eqnarray}
E_{ij}&=&\delta_{ij}   -\delta_{ij}\hat{c}_{00}   -\delta_{ik}\delta_{jl}(c_{kl}+c_{lk})  ,
\end{eqnarray}
\begin{eqnarray}
F_{ijk}&=&\delta_{ij} b_k - \delta_{jk} b_i  -m \delta_{jk}\hat{d}_{i0} \nonumber
\\
&&+ \frac{1}{2}m\epsilon_{klm}\delta_{ij}\hat{g}_{lm0} - \frac{1 }{2} \epsilon_{ilm} \delta_{jk}\left(m\hat{g}_{lm0}+H_{lm} \right),
\end{eqnarray}
\begin{eqnarray}
G_{ijk}&=&m \delta_{jk}(\hat{d}_{0i}+\hat{d}_{i0})  +m \epsilon_{ikl}\hat{g}_{0lj} -m \epsilon_{ikl}\hat{g}_{lj0}.
\end{eqnarray}
In these last expressions $m$ is the particle's mass, $p_i$ are the components of the momentum operator, $\sigma_i$ are the Pauli matrices, $\epsilon_{ijk}$ is the totally antisymmetric tensor with $\epsilon_{123}=1$, and all the indices run from $1$ to $3$ with the convention of summing over repeated indices. The SME coefficients for the sector we study are $a_\mu$, $b_\mu$, $c_{\mu\nu}$, $d_{\mu\nu}$, $e_\mu$, $f_\mu$ $g_{\mu\nu\rho}=-g_{\nu\mu\rho}$ and $H_{\mu\nu}=-H_{\nu\mu}$, with Greek indices running from $0$ to $3$. In the case we analyze, the SME coefficients get `redshifted' by a factor $(1+\Phi)^{-n}$, where $n$ is the number of zero indices in the corresponding coefficient. Thus, to get a compact expression we introduce a caret on top of some of the SME coefficients with the convention that a coefficient with a caret has been redshifted. For example, $\hat{b}_0\equiv b_0(1+\Phi)^{-1}$ and $\hat{g}_{0i0}\equiv g_{0i0}(1+\Phi)^{-2}$.

The nonrelativistic hamiltonian we have calculated agrees with that of Ref.\ \refcite{Hehl} when the SME coefficients are set to zero and it coincides with the result of Ref.\ \refcite{KostLane} in the limit when $\Phi=0$. In addition, where there is overlap, it agrees with the nonrelativistic hamiltonian presented in Ref.\ \refcite{KostTasson}. Even though $p_i$ acts on the SME coefficients when taking the adjoint, it is possible to verify that our hamiltonian is hermitian. Moreover, $a_\mu$ and $e_\mu$ only appear in the combination $a_\mu-me_\mu$ and the antisymmetric part of $d_{\mu\nu}$ only shows up in $H_{\mu\nu}+m {\varepsilon_{\mu\nu}}^{\rho \sigma}d_{\rho\sigma}/2$ where $\varepsilon_{\mu\nu\rho\sigma}$ is the spacetime volume form. Also, the antisymmetric part of $c_{\mu\nu}$ and $f_\mu$ do not enter into our hamiltonian. These facts are consistent with the analysis sketched in Ref.\ \refcite{Kost} regarding the freedom to redefine the fermionic field. 

Note that our hamiltonian is not invariant if we add a constant to $\Phi$. However, it is possible to see that, in the coordinates we work, the point where $\Phi = 0$ has physical meaning: it is where the observer/detector is located. Thus, adding a constant to $\Phi$ amounts to relocating the detector at a different height and, in order to verify that the physics is invariant under such transformation, one needs to take into the account the redshift on the detector. In addition, the gravitational effects in the hamiltonian \refeq{bonder:Hamiltonian} are only relevant for experiments where the particles probe regions with different $\Phi$. Thus, the best candidates to search for the effects of this hamiltonian are interferometry experiments like those of Refs. \refcite{Muller}. Also, the nonrelativistic hamiltonian for antiparticles has been calculated and it can be obtained from equation \refeq{bonder:Hamiltonian} with the replacements given in Ref.\ \refcite{KostLane} that link the particle and 
antiparticle hamiltonian in the nongravitational case. Furthermore, since each component of the SME coefficients gets redshifted in a different way, our result suggests that by doing experiments at several heights, it should be possible to disentangle the bounds that are usually placed on the linear combinations of these coefficients.

To summarize, we have obtained the nonrelativistic hamiltonian for the minimal matter SME sector in the presence of a uniform newtonian gravitational potential. The spin-dependent terms of this hamiltonian are presented here for the first time, and may lead to new experiments that will allow us to keep testing Lorentz invariance.

\section*{Acknowledgments}
I wish to thank Alan Kosteleck\'y for many helpful discussions. This work was supported by the Department of Energy under grant DE-FG02-13ER42002 and by the Indiana University Center for Spacetime Symmetries.

\end{document}